\documentclass[amsmath,amssymb,pra,final,showpacs,twocolumn]{revtex4-1}
 
\usepackage{bm,hyphenat,xspace}
\usepackage{graphicx,epsfig}

\newcommand {\mbf}[1]{{\mathbf{#1}}}

\newcommand{\A}[2]{{}^{#1}\mathrm{#2}}

\begin{document}

\title {Comparison of Pauli projection and supersymetric transformation methods
  for three-body nuclear structure and reactions}
  
\author{A.~Deltuva} 
\affiliation
{Institute of Theoretical Physics and Astronomy, 
Vilnius University, Saul\.etekio al. 3, LT-10257 Vilnius, Lithuania
}

\received{November 26, 2025} 

\begin{abstract}
  Three-body Faddeev-type equations for bound, resonant, and scattering states
  in systems with a nuclear core and two nucleons are solved using
  the momentum-space framework. Two approaches for eliminating the
  Pauli-forbidden deeply-bound states are compared:
  projecting out those states by a nonlocal term in the potential,
  and by using a supersymmetric transformation of the potential. While
  the former method is preferred by the experimental data for 
  deuteron-$\A{4}{He}$ scattering, the results for bound and resonant states
  do not indicate a clear superiority of a single method. Instead,
  systematic differences between them are found.
\end{abstract}

 \maketitle

\section{Introduction \label{sec:intro}}

Understanding  nuclear structure and reactions in terms of simplified few-body
models is among the major goals of nuclear physics. This introduces effective degrees
of freedom such as clusters of $\alpha$ particles or a nuclear core plus external
nucleons, whose interactions are described by effective two-body (and possibly three-body) potentials,
with parameters fitted to binding or low-energy scattering data.
The fermionic character of a many-body nuclear system plays a crucial
role, demanding to impose  antisymmetry constraints on the
nuclear wave functions. The deepest bound states supported by those effective potentials
correspond to the states  implicitly occupied by the internal core nucleons, and thus are Pauli forbidden
for external nucleons.
In simple approaches to nuclear reactions such as the distorted-wave Born approximation (DWBA) 
and adiabatic distorted-wave approximation (ADWA) \cite{timofeyuk:20a} only the core plus nucleon
bound-state wave functions are needed for the assumed  energy level.
The problem of Pauli-forbidden states becomes more serious in full three-body calculations,
where the external nucleons cannot occupy those deep states embedded into three-body space.
Without affecting higher-lying (Pauli allowed)
and continuum states, this restriction can be implemented in several ways:
(i) The Pauli-forbidden states are projected out from the considered model space;
this can be achieved by adding a strong repulsive nonlocal projection term to the two-body potential
\cite{kukulin:84a,schellingerhout:93a,thompson:00}
or, in the adiabatic expansion method \cite{garrido:97a,garrido:99a},
omitting the subset of states corresponding to Pauli-forbidden states.
(ii) The supersymmetric (SS) transformation is applied
to the potential \cite{baye:susy}  creating a repulsive $r^{-2}$ term,
that eliminates the deepest states but keeps all the higher states unchanged.
A less sophisticated  variation of (ii) is  a potential with repulsive core (RC) \cite{thompson:00},
which, however, does not preserve exact phase equivalence to the original potential.

The Pauli projection (PP) and SS transformation methods have been compared in several
works. Hesse et al. \cite{hesse:99a} calculated features of weakly bound two-neutron halo nuclei
$\A{6}{He}$, $\A{11}{Li}$, and $\A{14}{Be}$. They found that SS type potentials yield more binding, but
after the potential rescaling to fit the same binding energy the radii obtained with PP and SS methods
turn out to be very close.  Garrido et al. \cite{garrido:99a} studied $\A{6}{He}$ and $\A{11}{Li}$ and reached similar conclusions.
Thompson et al. \cite{thompson:00} considered $\A{6}{He}$ and $\A{6}{Be}$ systems, revealing
no significant differences between several treatments of Pauli forbidden states in
bound and resonant states, but visible effects for nonresonant continuum.
For example, the dipole distribution for $\A{6}{He}$ calculated in the SS method was systematically higher
than in the PP method, though the shape was similar.

The most detailed study of the continuum is offered by exclusive scattering processes. The continuum of
the $\A{4}{He}$ core and two nucleons was considered in Ref.~\cite{deltuva:06b} for the
examples of  deuteron-$\A{4}{He}$ elastic scattering and breakup, revealing very large differences
between the PP and RC models for the nucleon-$\A{4}{He}$ potential, not only in magnitude but also in shape,
with the experimental data clearly preferring the PP model.
As the SS method was not implemented,  it remained unclear whether the findings were not due to
the violation of the phase equivalence. This question will be sorted out in the present work
using rigorous Faddeev-type scattering equations for transition operators
\cite{alt:67a,deltuva:06b}, solved in the momentum space.
Note that due to high excitation energy of $\A{4}{He}$ this is one of few
 systems involving composite nuclei
 that is described not by optical but by real potentials in the continuum.
 The complex potentials do not support real bound states, and therefore
 are excluded from the discussion.

Regarding the bound states, the comparison of the PP and SS models will be performed not only
for weakly bound two-neutron halo nuclei such as $\A{11}{Li}$ and $\A{19}{B}$
but also for more tightly bound systems accomodating
bound excited states such as $\A{16}{C}$ and $\A{18}{O}$. The aim is not to fit their binding energies but
to study the emergence of differences
between the PP and SS and possibly find a systematic pattern.

Finally, the unbound $\A{16}{Be}$ nucleus will be studied. This will demonstrate the ability
of the momentum-space transition operator method to determine the resonance properties
for core plus two-neutron systems. Furthermore, it appears that existing $\A{16}{Be}$
calculations \cite{lovell:16be,casal:19a} performed in the coordinate space
rely on the SS model, presumably taking advantage of 
its local form, in contrast to the nonlocal PP potentials. In the momentum-space integral equation
framework both local and nonlocal potentials are treated on the same footing.

Section II briefly recalls the description of three-body bound and scattering states using
transition operators together with the essential aspects of the calculations.
Section III presents results scattering observables, bound state properties, and
resonances.
Discussion and conclusions are collected in Sec. IV.

\section{Three-body bound-state and continuum equations \label{sec:eq}}

Faddeev equations \cite{faddeev:60a} employ the decomposition of the three-body wave function
$|\Psi\rangle = \sum_{\alpha=1}^3 |\psi_\alpha\rangle$ into the  amplitudes 
$ |\psi_\alpha\rangle$. For the three-body bound state with energy $E_B < 0$ (measured
from the threshold of three free particles) 
the Faddeev amplitudes obey the system of coupled equations
\begin{equation} \label{eq:phi}
| \psi_{\alpha} \rangle = G_0(E_B) \sum_{\beta=1}^3 \bar{\delta}_{\alpha\beta} T_\beta(E_B)
| \psi_{\beta} \rangle.
\end{equation}
The Greek subscripts label the spectator particle in the odd-man-out notation,
$\bar{\delta}_{\alpha\beta} = 1-{\delta}_{\alpha\beta}$,
\begin{equation} \label{eq:G0}
G_0(Z) = (Z-H_0)^{-1}
\end{equation}
is the free resolvent with the kinetic energy operator $H_0$, and
\begin{equation} \label{eq:t2b}
T_\alpha(Z) = v_\alpha + v_\alpha G_0(Z) T_\alpha(Z)
\end{equation}
is the two-particle transition operator acting in the three-body space,
with $v_\alpha$ being the corresponding two-particle potential.

The transition operators $U_{\beta\alpha}(Z)$ in the Alt-Grassberger-Sandhas version \cite{alt:67a}
obey the system
\begin{equation}  \label{eq:Uba}
U_{\beta\alpha}(Z)  = \bar{\delta}_{\beta\alpha} \, [G_{0}(Z)]^{-1}  +
\sum_{\eta=1}^3    \bar{\delta}_{\beta\eta} \, T_{\eta}(Z)  \,
G_{0}(Z) U_{\eta\alpha}(Z).
\end{equation}
The physical amplitudes for reactions at energy $E$ are obtained as on-shell matrix elements
of the transition operators $U_{\beta\alpha}(E+i0)$; see Ref.~\cite{deltuva:06b} for the
definitions of deuteron-$\A{4}{He}$ elastic and  breakup amplitudes.
At $Z=E_B$ the transition operators $U_{\beta\alpha}(E_B)$ have the bound-state pole, while
three-body resonances, if existing, correspond to poles of $U_{\beta\alpha}(Z_R)$ at
$Z_R = E_R - i\Gamma/2$ in the second Riemann sheet of the complex energy plane, with
the resonance position $E_R$ and width $\Gamma$. If the resonance is not very broad,
those parameters can be determined from the energy dependence of  matrix elements obtained
at real physical energies as done in the fictitious three-neutron system with
enhanced interactions \cite{deltuva:18a}, i.e.,
\begin{equation}  \label{eq:Ures}
  U_{\beta\alpha}(E+i0)  \approx
  \frac{\tilde{U}_{\beta\alpha}^{(-1)}}{E-E_R + i\Gamma/2} +
  \tilde{U}_{\beta\alpha}^{(0)} + \ldots.
\end{equation}

Equations (\ref{eq:phi}) and (\ref{eq:Uba}) are
 solved  in the momentum-space partial-wave representation. Three  sets of basis functions
$|p_\alpha q_\alpha 
(l_\alpha \{ [ L_\alpha (s_\beta s_\gamma)S_\alpha] j_\alpha s_\alpha\} \mathcal
 {S}_\alpha) J M \rangle $
 are used with $(\alpha,\beta,\gamma)$ being cyclic permutations of $(1,2,3)$.
The  Jacobi momenta $p_\alpha$ and  $q_\alpha$ refer to the 
pair $(\beta,\gamma)$ and spectator $\alpha$, their associated orbital
angular momenta are $L_\alpha$ and  $l_\alpha$. Together with the particle spins
$s_\alpha,s_\beta, s_\gamma$  they are coupled
to the total angular momentum $J$, $M$ being its projection.
$S_\alpha$ and  $j_\alpha$ are the spin and total angular momentum of the pair $\alpha$, and
$ \mathcal{S}_\alpha$ is the three-body channel spin.
In the core plus two-neutron systems  including partial waves with $L_\alpha \le 3$ leads to well converged results.
However, if a proton is involved, such as in the deuteron-$\A{4}{He}$ scattering, the screening
and renormalization method \cite{taylor:74a,alt:80a,deltuva:05a}
for the inclusion of the Coulomb force requires screening radii up to 20 fm and, consequently,
$L_\alpha$ up to 15 for the proton-$\A{4}{He}$ pair.

\section{Results  \label{sec:res}}

The two-nucleon interaction is taken to be the high-precision charge-dependent Bonn (CD Bonn) potential \cite{machleidt:01a},
though any realistic two-nucleon potential leads to a very similar results as proven in a number of previous works,
e.g., \cite{deltuva:06b,deltuva:09d}.
The nucleon-nucleus potential is assumed to be of the standard Woods-Saxon form with the
spin-orbit term, in the coordinate space given by
\begin{equation}  \label{eq:VnA}
v(r) = -V_c f(r,R,a) + \mbf{s_N} \cdot \mbf{L} \,
V_{so} \, \frac{1}{r} \frac{d}{dr}f(r,R,a),
\end{equation}
with $f(r,R,a) = [1+\exp((r-R)/a)]^{-1}$, the diffuseness $a$, and radius $R = r_0 A^{1/3}$.
The strengths $V_c$ and $V_{so}$ are adjusted case by case and may be partial-wave dependent;
for brevity the spectator label $\alpha$ will be suppressed in  further notation.
The interaction is assumed to be independent of the core spin.

If the potential (\ref{eq:VnA}) supports a Pauli-forbidden state $|\phi_P^{L j}\rangle$,
projecting it out is equivalent to moving
to infinetely high energy \cite{schellingerhout:93a} via modified potential containing a
nonlocal projection term,
\begin{equation}  \label{eq:VnAP}
v^P = v + |\phi_P^{L j}\rangle \Gamma_P \langle \phi_P^{L j} | .
\end{equation}
Formally $\Gamma_P \to \infty$, but practically that limit is reached with $\Gamma_P$ ranging from 1 to 10 GeV,
where the  predictions, to  good accuracy, become independent of  $\Gamma_P$.

In the SS method the potential acquires a local repulsive correction \cite{baye:susy} that
at short distances behaves like $r^{-2}$ but, except for the Pauli-forbidden state, has the same spectrum as the
original potential. The singular term of the SS potential becomes harmless 
after transformation to the momentum space.

\subsection{Deuteron-$\A{4}{He}$ scattering}

\begin{figure}[!]
\begin{center}
\includegraphics[scale=0.7]{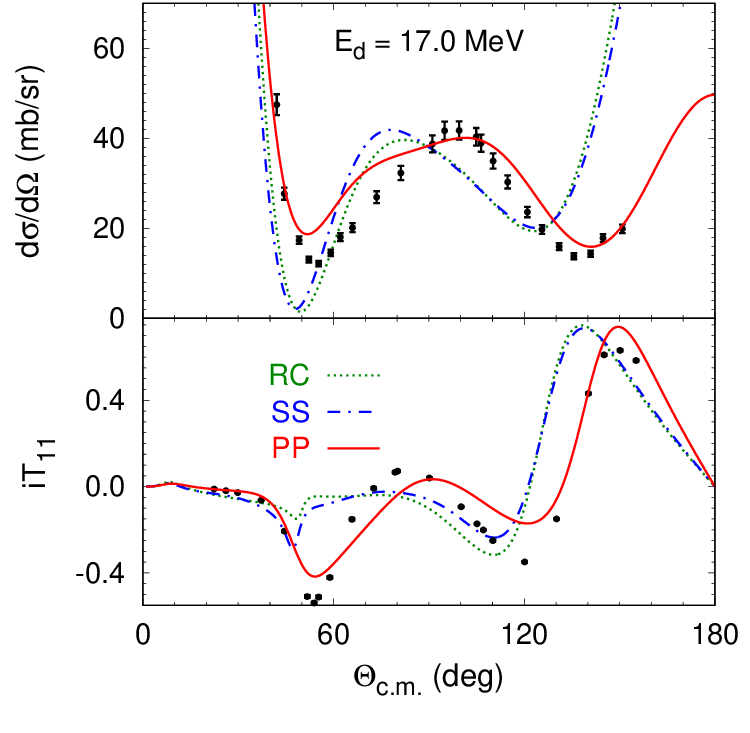}
\end{center}
\caption{\label{fig:elastic} (Color online) 
  The differential cross section and deuteron vector analyzing power $iT_{11}$
for the deuteron-$\A{4}{He}$ elastic scattering  at 
17~MeV deuteron beam energy as functions of the c.m. scattering angle.
Results of PP (solid  curves), SS (dashed-dotted curves),
and RC (dotted curves) models  are compared. The experimental
data are from Ref.~\cite{gruebler:79a}.}
\end{figure}

\begin{figure}[!]
\begin{center}
\includegraphics[scale=0.7]{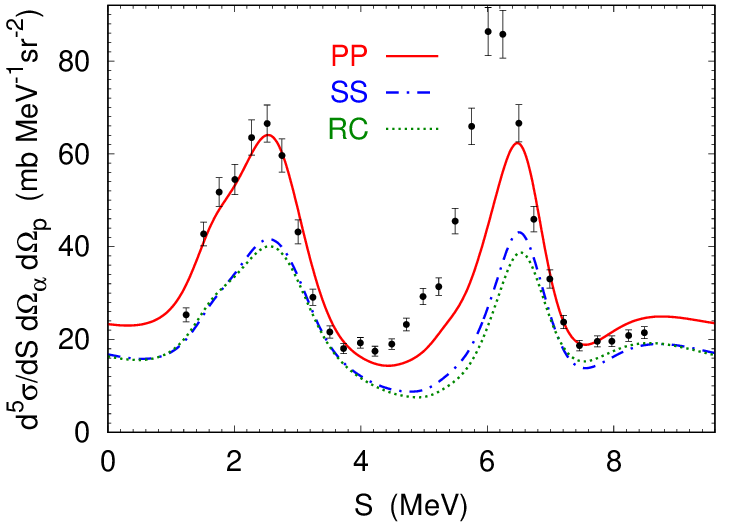}
\end{center}
\caption{\label{fig:breakup} (Color online) 
The fully exclusive differential cross section for the deuteron breakup
in collision with the 15~MeV  $\A{4}{He}$ nucleus. 
It is shown as a function of the arclength $S$ along the kinematical curve
in a coplanar kinematical configuration with 17.1 and 50.5 deg polar angles  of
$\A{4}{He}$ and proton, respectively, and with 180 deg difference in their azimuthal angles.
Curves are as in Fig.~\ref{fig:elastic}, and
the experimental data are taken from Ref.~\cite{koersner:77}.}
\end{figure}

\begin{figure}[!]
\begin{center}
\includegraphics[scale=0.7]{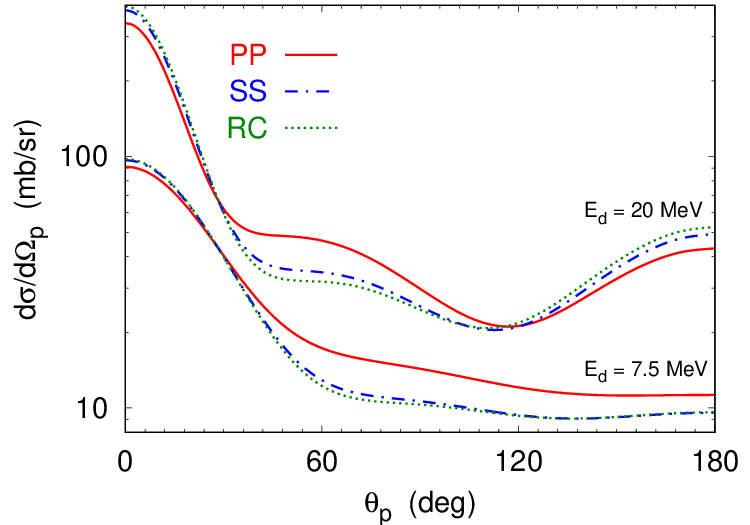}
\end{center}
\caption{\label{fig:d2s} (Color online) 
  The semi-inclusive differential cross section for the $\A{4}{He}(d,p)$ reaction
at 7.5 and 20 MeV deuteron beam energies as a function of the proton  scattering angle in the c.m. frame.
Curves are as in Fig.~\ref{fig:elastic}.}
\end{figure}

The deuteron-$\A{4}{He}$ elastic scattering and breakup reactions have been studied in Ref.~\cite{deltuva:06b}.
The comparison of the PP and RC models in Ref.~\cite{deltuva:06b}
clearly revealed the superiority of the former in reproducing the
experimental data. Though both SS and RC potentials are local,
 the RC potential remains finite at the origin while the SS one has a singularity, thus, it is not
 clear to what extent their predictions will differ, and whether SS will be able to remedy the deffects
 of RC. Figures \ref{fig:elastic} - \ref{fig:d2s}
 present the comparison of predictions
 using PP, SS, and RC potentials, with their parameters given in Ref.~\cite{thompson:00},
 labeled as PP, PS, and PC models there, respectively.
 They are defined for $L \le 2$ partial waves.
 All the shown examples --- i.e.,  the differential cross section and the deuteron vector analyzing power in the
 deuteron-$\A{4}{He}$ elastic scattering  at 17~MeV deuteron beam energy
 in Fig.~\ref{fig:elastic}, the fully exclusive fivefold differential cross section for the
 deuteron breakup at 15 MeV $\A{4}{He}$ beam energy in Fig.~\ref{fig:breakup}, and
 the differential cross section for the  semi-inclusive breakup reaction $\A{4}{He}(d,p)$
 in Fig.~\ref{fig:d2s} --- deliver the same message: the predictions of RC and SS models are quite close
 to each other  and clearly different from the PP ones. The available experimental data
 \cite{gruebler:79a,koersner:77} favor the PP model, as already found in Ref.~\cite{deltuva:06b},
 where more examples are given. There are no data available for the
 semi-inclusive  reaction $\A{4}{He}(d,p)$, but it serves as a global characteristic for the breakup
 in the whole phase space. It has quite a significant contribution from the $\A{5}{He}$ 
 resonant  state $1p_{3/2}$.
 
The difference between the PP and SS/RC models to a lesser extent shows up also in the binding energy of $\A{6}{Li}$,
with the predictions (relative to the core plus nucleons threshold) being 3.48, 3.66, and 3.65 MeV, respectively,
quite close to the experimental value of 3.69 MeV.

Some potential models project out not the deep state wave function but 
the ground state of the harmonic oscillator (HO) \cite{niessen:92}.
As discussed in  Ref.~\cite{deltuva:06b} this does not lead to a significant
 difference as compared to the PP method. 
Kamada {\it et al.} \cite{kamada:00a} 
calculated the forbidden (called ghost) state 
microscopically and found that it can be approximated quite well by the 
HO function with the shell model mode $\Omega=0.245\, \rm{fm}^{-2}$.
The potential used in this work prefers a smaller value of
$\Omega \approx 0.21\, \rm{fm}^{-2}$, in a good agreement with Ref.~\cite{niessen:92}.

\subsection{Bound states \label{sec:bs}}

Systems consisting of a nuclear core and  two neutrons  have been extensively studied in the past
using three-body models, e.g., in
Refs.~\cite{thompson:00,hesse:99a,thompson:11li,nunes:96a,face:04,jensen:04a,deltuva:17c,hiyama:b19,o18-chin:25a}.
The used neutron-core interaction models range from PP to SS and RC, though quite often just a single
model was chosen. In this section the comparison of PP and SS models will be presented for several
nuclei, ranging from weakly to tightly bound. The potential parameters in
Eq.~(\ref{eq:VnA}) are taken from previous works where available.
All potentials support Pauli-forbidden  $L=0$ and 1 states, i.e., $1s_{1/2}$ and $1p_j$,
and no such states for higher $L$.

The case of $n$-$\A{9}{Li}$ is exceptional in the sense that only $1p_{3/2}$ is Pauli forbidden,
while its $1p_{1/2}$ state is not Pauli-forbidden but resonant.
For $n$-$\A{9}{Li}$ the model P4 of Ref.~\cite{thompson:11li} is taken.
It has a virtual $2s_{1/2}$ state. The parameters were tuned in Ref.~\cite{thompson:11li}
to reproduce $\A{11}{Li}$ binding energy of 0.3 MeV in the PP treatment.

For $n$-$\A{17}{B}$ the spin-orbit force and $L$-dependence are neglected as in Ref.~\cite{hiyama:b19},
while $V_c = 42.232$ MeV, $R=3$ fm, $a=0.6$ fm, reproducing a virtual  $2s_{1/2}$ state with
large negative two-body scattering length
of -100 fm \cite{hiyama:b19}.

For $n$-$\A{18}{C}$ the parameters are taken from the DPp model Ref.~\cite{deltuva:17c},
supporting a  bound $2s_{1/2}$ state with $E_B=-0.58$ MeV. The potential has no spin-orbit force and
uses the same parameters in all waves.
  
For $n$-$\A{14}{C}$ the parameters are taken from Ref.~\cite{deltuva:09d}, adjusted to the low-energy
spectrum of $\A{15}{C}$ and neutron separation energy of $\A{14}{C}$.

For $n$-$\A{16}{O}$ the parameters are taken from Ref.~\cite{o18-chin:25a}, again adjusted
to the low-energy spectrum of $\A{17}{O}$.
Since they are defined for $L \le 2$ partial waves,
the results presented in tables will use  the restriction $L \le 2$, in contrast to
$L \le 3$ for other nuclei.

In the considered cases the binding energies of the Pauli-forbidden $1s_{1/2}$ states range from
22 MeV ($n$-$\A{17}{B}$) to 37 MeV ($n$-$\A{16}{O}$),
and from 4 MeV ($n$-$\A{9}{Li}$) to 32 MeV ($n$-$\A{16}{O}$) in  $1p_j$ states.

The calculated ground-state binding energies are collected in Table \ref{tab:gs}. The numerical accuracy
is 1 ($\A{19}{B}$) to 10 keV ($\A{18}{O}$). The predictions are in good
agreement with previous calculations where available
\cite{thompson:11li,hiyama:b19,o18-chin:25a},  the remaining discrepancies most likely being due to
differences in the included  partial waves, two-neutron potential, and even parameter value roundoff.
The experimental values from Ref.~\cite{nndc} are listed as well.
The aim is not to fit these values, but they turn out to be quite close to one of the models
used to remove Pauli-forbidden states. However, the most important  message from Table \ref{tab:gs}
is that  the SS model systematically predicts larger binding energies as compared to PP.

\begin{table}[!]
  \caption{Ground-state binding energies $|E_B|$ (in MeV) for the considered nuclei calculated
    using PP and SS models.
    The last line $\A{18}{O}$ results are obtained including also
    $L=3$ waves with the same parameters as in $L=2$ waves.
The experimental data are from Ref.~\cite{nndc}.
The experimental value for $\A{19}{B}$ is a compilation from several measurements 
and theoretical evaluations, having a large uncertainty of about
0.4 MeV; see also Ref.~\cite{hiyama:b19}.
}
\label{tab:gs}
\centering
\begin{ruledtabular}
\begin{tabular}{{l}*{3}{r}}
  & PP  & SS &  Expt. \\ \hline
$\A{19}{B}$  & 0.185 & 0.383 & 0.145 \\
$\A{11}{Li}$ & 0.296 & 0.855 & 0.369 \\
$\A{20}{C}$  & 2.675 & 3.644 & 3.560 \\
$\A{16}{C}$  & 4.342 & 5.299 & 5.468 \\
$\A{18}{O}$  & 11.458 & 12.773 &  12.188 \\
$\A{18}{O}$  & 11.550 & 12.917 &  12.188 \\
\end{tabular}
\end{ruledtabular}
\end{table}

\begin{figure}[!]
\begin{center}
\includegraphics[scale=0.66]{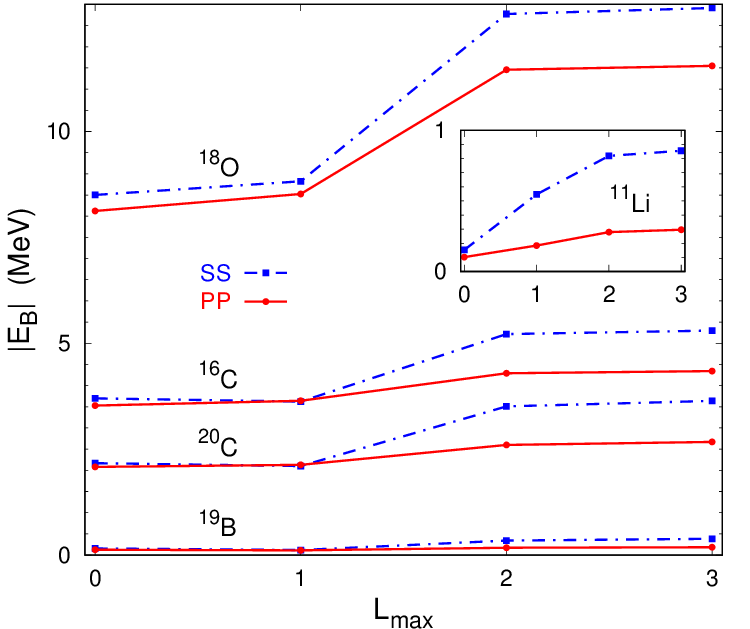}
\end{center}
\caption{\label{fig:Bgs} (Color online) 
  Ground-state binding energies of
  $\A{19}{B}$,  $\A{11}{Li}$, $\A{20}{C}$, $\A{16}{C}$, and $\A{18}{O}$ nuclei,
  calculated including different number of  angular momentum states.
The predictions of SS and PP models are compared.
The lines are for guiding the eye only.
}
\end{figure}

Notably, all of these core-neutron systems support a virtual or weakly bound $2s_{1/2}$ state.
Together with the virtual two-neutron ${}^1S_0$ state this fulfills the condition
for the manifestation of the Efimov physics \cite{jensen:04a}.
As a consequence, the three-body ground states emerge
already under restriction to  $L=0$ partial waves.
The universality predicts low sensitivity to the short-range details, and indeed the
binding energies calculated with solely $L=0$ partial waves in PP and SS models are very similar.
It is interesting to study how the differences between the PP  and SS aproaches emerge
starting from this roughly universal situation.
The results for three-body binding energy calculations
including the neutron-core interaction in partial waves with $L \le L_{\mathrm{max}}$
are shown in Fig.~\ref{fig:Bgs}. 
The two-neutron potential is taken up to $L=3$, but $L>0$ waves yield only small contribution
not affecting the picture qualitatively.
Except for $\A{11}{Li}$, all  $1p_j$ waves contain Pauli-forbidden states, the interaction is
predominantly Pauli-repulsive with little effect for the binding energy.
In contrast, $L=2$ core-neutron waves contain no Pauli-forbidden states and yield
quite appreciable contribution. Most importantly, in the SS model it is considerably
larger than in PP, and this is decisive for the SS-PP difference.
The changes due to $L=3$ waves are small.
In the $\A{11}{Li}$ case, shown in the inset of  Fig.~\ref{fig:Bgs}, the
$1p_{1/2}$ state has no Pauli repulsion and therefore already $L_{\mathrm{max}}=1$
results show an increase of the binding energy, further enhanced by $L=2$ waves.
Again, in the SS model the  rise of the binding energy is considerably larger than in PP.

\begin{figure}[!]
\begin{center}
\includegraphics[scale=0.66]{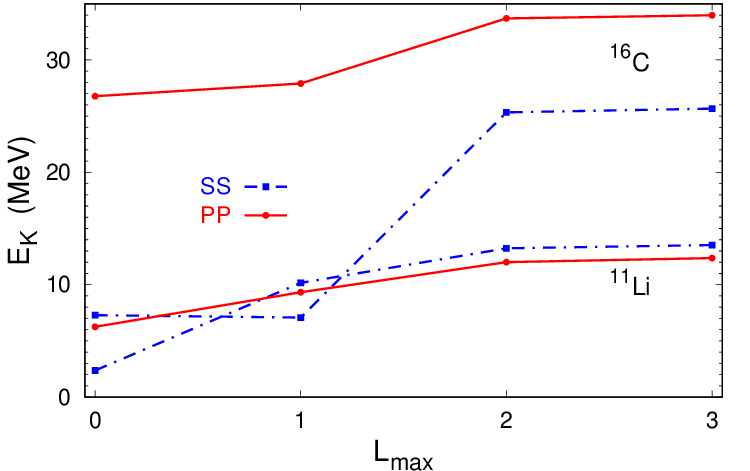}
\end{center}
\caption{\label{fig:Kgs} (Color online) 
  Kinetic energy expectation value for ground state of $\A{11}{Li}$ and $\A{16}{C}$ nuclei,
  calculated including different number of angular momentum states.
The predictions of SS and PP models are compared.
The lines are for guiding the eye only.
}
\end{figure}

Thus, Fig.~\ref{fig:Bgs} indicates that three-body ground state energies calculated
using SS and PP models are close as long as only low partial waves containing
Pauli-forbidden states are included. However, the interplay of lower and higher partial waves
leads to the binding energy increase that is significantly larger for SS.
The reason might be different wave functions, i.e., different weights of various momentum
components. A global measure for the low/high momentum components is the expectation value
of the kinetic energy $E_K = \langle \Psi | H_0 | \Psi \rangle$.
It is presented in Fig.~\ref{fig:Kgs} for  $\A{11}{Li}$ and  $\A{16}{C}$; the remaining cases
are qualitatively similar to  $\A{16}{C}$. It reveals drastic differences already in
low partial waves, the kinetic energy in the PP model being considerably higher than in SS.
When higher partial waves without Pauli repulsion are included, $E_K$ rises more strongly
in the SS model. For  $\A{11}{Li}$ this happens at  $L_{\mathrm{max}}=1$ while for the remaining
cases at $L_{\mathrm{max}}=2$. 
The difference in $E_K$ by far exceeds the one in binding energy, indicating
that the expectation value of the potential energy $E_V = E_B - E_K$
in the PP model is also higher in absolute value despite that the binding energy
is lower, only  $\A{11}{Li}$ being the exception.
Thus, the observed  difference in binding energies is a result of partial cancellations
between kinetic and potential energies. A conjecture is that the PP model, explicitly
imposing ortogonality to the Pauli-forbidden state, induces higher momentum components,
leading to larger $E_K$. On the other hand, the repulsive core of the SS model leads to
lower $|E_V|$ and possibly increases the weight of larger distance components  in low partial waves,
thereby making the coupling to higher waves more efficient, and increasing their
effect both for $E_K$ and $E_B$ as observed in Figs.~\ref{fig:Bgs} and \ref{fig:Kgs}.
For example, in $\A{16}{C}$ the weights of the dominating $L=0$ and $L=2$ core-neutron components
are around 62\% and 35\% in PP but they are 51\% and 46\% in SS model, respectively. The remaining 3\% are distributed
among $L=1$ and $L=3$  waves.

\begin{figure}[!]
\begin{center}
\includegraphics[scale=0.66]{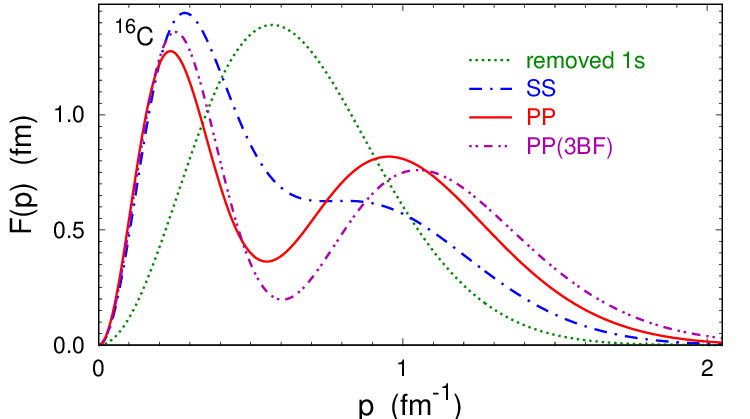}
\end{center}
\caption{\label{fig:md16c} (Color online) 
 Core-neutron relative momentum distribution  in the ground state of  $\A{16}{C}$ 
and in the Pauli-forbidden $1s_{1/2}$ state of $\A{15}{C}$ (dotted curve).
Results of SS (dashed-dotted curve),  PP (solid  curve), 
and PP+3BF (double-dotted-dashed curve) models  are compared.
}
\end{figure}

The core-neutron relative momentum distributions are compared in Fig.~\ref{fig:md16c}
for the example of $\A{16}{C}$, and turn out to be quite different.
The normalization is $\int_0^\infty F(p) dp = 1$.
In order to eliminate the possible effect due to the binding energy mismatch,  a phenomenological
three-body force (3BF) is added to the PP Hamiltonian; its form and resulting equations can be found in
Ref.~\cite{deltuva:13b}. With parameters $\Lambda=2\,\mathrm{fm}^{-1}$ and
$w_3=-6.2\,\mathrm{MeV fm}^6$ in the convention of Ref.~\cite{deltuva:13b} the PP+3BF model binding energy
coindides with the one of SS. The resulting momentum distribution is shifted towards larger momenta but
nevertheless remains closer to the one by PP. For curiosity, the momentum distribution of the
Pauli forbidden $1s_{1/2}$ state is shown as well. Its maximum almost coincides with local minimum in
PP and PP+3BF models, consistently with a vanishing overlap $\langle \phi_P^{L j} | \Psi \rangle$.
This does not take place in the SS model, where the weight of this overlap is 0.166, the momentum distribution
does not show a deep minimum and has considerably lower high-momentum tail.

Except for very weakly bound $\A{11}{Li}$ and $\A{19}{B}$ systems, the other considered nuclei
 support  excited states, at least those with total spin and parity $J^\Pi = 2^+$ and $0^+$; the later
will be labeled $0_2^+$ in order to distinguish from the ground state. For $\A{16}{C}$ and  $\A{18}{O}$  few more
excited states are available. The predictions for their binding energies are collected in Table \ref{tab:xs}. Interaction in partial waves with $L=2$
is necessary for these states to appear. Their binding energies again show a
systematic behavior, in the SS model being larger for $2^+$ states
by roughly 0.8 MeV but smaller by up to 0.7 MeV
for $0_2^+$ states as compared to PP.
Notably, both models predict a very weakly bound $\A{20}{C}(0_2^+)$
state, 50 to 90 keV below the $n$-$\A{19}{C}$ threshold;
this state is not known experimentally. For definite conclusions a more refined
$n$-$\A{18}{C}$ potential, especially in $L>0$ waves,
possibly including also the core excitation,
is needed, which is beyond the goals of the present study.
For further $\A{16}{C}$ and $\A{18}{O}$ states
the SS binding energies are slightly larger, but the
difference is not significant. This reduced sensitivity probably
can be explained by the fact that for higher $J$ values the partial waves
containing Pauli-forbidden states must couple with larger spectator orbital
momentum $l$, and their weight in the bound-state wave function is  small.
For example, in the $\A{18}{O}(4^+)$ state the weight of $L\le 1$ waves
is well below 0.5\%.
Regarding the experimental data, there is no clear preference of
any of the models. The predicted binding energies are roughly consistent with
the data for  $\A{18}{O}$, while for higher $\A{16}{C}$ states they are
slightly overestimated, the experimental data being closer to
the $\A{15}{C}$ threshold located at 1.218 MeV.

\begin{table}[!]
  \caption{Binding energies  $|E_B|$ (in MeV) for excited states of the
    considered nuclei calculated using PP and SS models. 
    The experimental data are from Ref.~\cite{nndc}.
    The $\A{18}{O}(0_2^+)$ in the theoretical model corresponds to the
    experimental  $\A{18}{O}(0_3^+)$ state, as explained in
    Ref.~\cite{o18-chin:25a}.
}
\label{tab:xs}
\centering
\begin{ruledtabular}
\begin{tabular}{{l}*{3}{r}}
  & PP  & SS &  Expt. \\ \hline
$\A{20}{C}(2^+)$  & 1.433 & 2.323 & 1.942 \\
$\A{16}{C}(2^+)$  & 3.337 & 4.058 & 3.702 \\
$\A{18}{O}(2^+)$  & 10.281 & 11.078 & 10.206 \\ \hline
$\A{20}{C}(0_2^+)$  & 0.668 & 0.632 &  \\
$\A{16}{C}(0_2^+)$  & 2.490 & 1.918 & 2.441 \\
$\A{18}{O}(0_2^+)$  & 7.823 & 7.133 & 6.852 \\ \hline
$\A{16}{C}(2_2^+)$  & 1.614 & 1.621 & 1.482 \\
$\A{18}{O}(2_2^+)$  & 8.456 & 8.601 & 8.268 \\ \hline
 $\A{16}{C}(3^+)$  & 1.617 & 1.706 & 1.380 \\
  $\A{18}{O}(3^+)$  & 7.429 & 7.605 & 6.811 \\ \hline
  $\A{16}{C}(4^+)$  & 1.668 & 1.716 & 1.326 \\
  $\A{18}{O}(4^+)$  & 8.624 & 8.810 & 8.633 \\
\end{tabular}
\end{ruledtabular}
\end{table}

Regarding the scattering calculations, 
the above neutron-nucleus potentials are only suitable
at very-low energies since, being real, they do
not account for inelastic processes. Here they are applied
for the neutron scattering off a core plus neutron system at zero
energy. The predicted values for  scattering lengths $a_{12}(J^\Pi)$
are collected in Table~\ref{tab:a}. In channels with higher $J$ values
the spins of the neutrons are aligned, producing the Pauli repulsion
at the level of external neutrons with very little sensitivity
to the details of the interaction. In fact, in those channels
$a_{12}(J^\Pi)$  is very close to the neutron-core scattering
length as observed also in Ref.~\cite{deltuva:17c}. In channels with
lower $J$ values $a_{12}(J^\Pi)$  is affected by the state 
closest to the threshold, where the differences in binding energy
predictions of PP and SS models are reflected in the $a_{12}(J^\Pi)$
differences, with the lower binding energy corresponding to
the larger scattering length. In particular, 
a very shallow $\A{20}{C}(0_2^+)$ state leads to large values
of $a_{12}(0^+)$ in $n$-$\A{19}{C}$ scattering.

\begin{table}[!]
  \caption{Scattering length (in units of fm) for neutron scattering
    off different two-cluster  nuclei (in the ground state)
    in a given total angular momentum
    and parity state. The numerical accuracy is better than 1\%.
}
\label{tab:a}
\centering
\begin{ruledtabular}
\begin{tabular}{{l}*{3}{r}}
   & $J^\Pi$ & PP  & SS  \\ \hline
  $n$-$\A{19}{C}$  & $0^+$ & 16.4 & 26.6 \\
  $n$-$\A{19}{C}$  & $1^+$ & 9.25 & 9.25 \\
  $n$-$\A{15}{C}$  & $0^+$ & 4.84 & 7.74  \\
  $n$-$\A{15}{C}$  & $1^+$ & 7.40 & 7.39  \\
  $n$-$\A{17}{O}$  & $2^+$ & 4.23 & 3.73  \\
  $n$-$\A{17}{O}$  & $3^+$ & 5.30 & 5.26  \\
\end{tabular}
\end{ruledtabular}
\end{table}

\subsection{$\A{16}{Be}$ resonances \label{sec:rsn}}

The $n$-$\A{14}{Be}$ potential parameters are taken from
Refs.~\cite{lovell:16be,casal:19a}, reproducing 
spectrum of $\A{15}{Be}$  $d$-wave resonances.
Partial waves with $L \le 3$ are included.
The $\A{16}{Be}$ resonance parameters are extracted using
Eq.~(\ref{eq:Ures}), about five terms are needed to achieve
the convergence of results.
The numerical accuracy is better than 0.02 MeV. The resonance
positions and widths are collected in Table \ref{tab:res}.

\begin{table}[!]
  \caption{$\A{16}{Be}$ resonance energies and widths (both in MeV)
     calculated using PP and SS models. 
    The experimental data are from Ref.~\cite{monteagudo:24a}.
}
\label{tab:res}
\centering
\begin{ruledtabular}
  \begin{tabular}{{l}*{6}{r}}   \multicolumn{1}{c}{}  &
    \multicolumn{2}{c}{ PP} & \multicolumn{2}{c}{ SS}
    & \multicolumn{2}{c}{Expt.}\\ 
  & $E_R$ & $\Gamma$ & $E_R$ & $\Gamma$ & $E_R$ & $\Gamma$ \\ \hline
$\A{16}{Be}(0^+)$  & 2.08 & 0.27 & 1.69 & 0.25 & 0.84 & 0.32 \\
$\A{16}{Be}(2^+)$  & 2.91 & 0.49 & 2.69 & 0.59 & 2.15 & 0.95\\
\end{tabular}
\end{ruledtabular}
\end{table}

For both $\A{16}{Be}$ states $0^+$ and $2^+$ the SS model predicts
the resonance at lower energy than PP. This is fully consistent
with the same $J^\Pi$ bound-state results, where the SS model
predicted more binding. As in Refs.~\cite{lovell:16be,casal:19a},
the chosen two-body potentials predict the resonances at higher
energies than experimental data, a possible remedy being an
attractive three-body force. On the other hand, except for a rather
well established $\A{15}{Be}(\frac52^+)$ resonance, the two-body
interaction is not well constrained by the data.

\section{Discussion and conclusions \label{sec:sum}}

Faddeev-type equations for bound and scattering states
in three-body  systems consisting of a nucleus and two nucleons
were solved in the momentum-space partial wave representation.
Two methods for eliminating the Pauli-forbidden states were compared,
either projecting those states out by a nonlocal projection term
in the two-body potential, or by a supersymmetric transformation
of the two-body potential. While some previous studies of this type
exist for weakly bound halo nuclei, this work presents a more systematic
study involving elastic and inelastic deuteron scattering, excited and
resonant states.

Results for the deuteron-$\A{4}{He}$ elastic scattering and breakup
clearly indicate that the PP method is favored by the experimental
data while SS and RC methods are close to each other and fail similarly.
For bound states and resonances there is no clear preference 
of the method, but systematic differences between them are established:
the SS method predicts larger binding energy (lower resonance position)
for the ground state and $2^+$ excited state, while the PP method
leads to larger binding energy for  $0_2^+$ excited states.
These differences in binding energy predictions  are reflected in the
respective scattering lengths for the neutron scattering off the
core plus neutron system.
The differences are small as long as only low partial waves containing
Pauli-forbidden states are included, but the interplay of lower
and higher partial waves
leads to a binding energy increase that is significantly larger for SS.
Furthermore, the expectation value for the kinetic energy is typically
higher in the PP method, with sizable differences in momentum
distributions as well. This could show up in breakup reactions,
where the amplitudes depend on the momentum distributions in the
bound state \cite{deltuva:13b}.
The higher excited states have quite small components with low partial
waves and therefore show only mild difference between PP and SS treatments.
Notably, in both methods a very weakly bound $\A{20}{C}(0_2^+)$
excited state emerges that is not yet known experimentally. 
A more definite conclusions regarding this (and some other) exotic nuclei
require better constrained nucleon-nucleus potentials.
\vspace{1mm}



\begin{thebibliography}{10}

\bibitem{timofeyuk:20a}
N.~K. Timofeyuk and R.~C. Johnson, Progress in Particle and Nuclear Physics
  {\bf 111},  103738  (2020).

\bibitem{kukulin:84a}
V. Kukulin, V. Krasnopolsky, V. abnd~Voronchev, and P. Sazonov, Nuclear Physics
  A {\bf 417},  128  (1984).

\bibitem{schellingerhout:93a}
N.~W. Schellingerhout, L.~P. Kok, S.~A. Coon, and R.~M. Adam, Phys. Rev. C {\bf
  48},  2714  (1993).

\bibitem{thompson:00}
I.~J. Thompson, B.~V. Danilin, V.~D. Efros, J.~S. Vaagen, J.~M. Bang, and M.~V.
  Zhukov, Phys. Rev. C {\bf 61},  024318  (2000).

\bibitem{garrido:97a}
E. Garrido, D. Fedorov, and A. Jensen, Nuclear Physics A {\bf 617},  153
  (1997).

\bibitem{garrido:99a}
E. Garrido, D. Fedorov, and A. Jensen, Nuclear Physics A {\bf 650},  247
  (1999).

\bibitem{baye:susy}
D. Baye, Phys. Rev. Lett. {\bf 58},  2738  (1987).

\bibitem{hesse:99a}
M. Hesse, D. Baye, and J.-M. Sparenberg, Physics Letters B {\bf 455},  1
  (1999).

\bibitem{deltuva:06b}
A. Deltuva, Phys.~Rev.~C {\bf 74},  064001  (2006).

\bibitem{alt:67a}
E.~O. Alt, P. Grassberger, and W. Sandhas, Nucl.~Phys. {\bf B2},  167  (1967).

\bibitem{lovell:16be}
A.~E. Lovell, F.~M. Nunes, and I.~J. Thompson, Phys. Rev. C {\bf 95},  034605
  (2017).

\bibitem{casal:19a}
J. Casal and J. G\'omez-Camacho, Phys. Rev. C {\bf 99},  014604  (2019).

\bibitem{faddeev:60a}
L.~D. Faddeev, Zh.~Eksp.~Teor.~Fiz. {\bf 39},  1459  (1960) [Sov.~Phys. JETP
  {\bf 12}, 1014 (1961)].

\bibitem{deltuva:18a}
A. Deltuva, Phys. Rev. C {\bf 97},  034001  (2018).

\bibitem{taylor:74a}
J.~R. Taylor, Nuovo Cimento B {\bf 23},  313  (1974); M.~D. Semon and J.~R.
  Taylor, Nuovo Cimento A {\bf 26}, 48 (1975).

\bibitem{alt:80a}
E.~O. Alt and W. Sandhas, Phys.~Rev.~C {\bf 21},  1733  (1980).

\bibitem{deltuva:05a}
A. Deltuva, A.~C. Fonseca, and P.~U. Sauer, Phys.~Rev.~C {\bf 71},  054005
  (2005).

\bibitem{machleidt:01a}
R. Machleidt, Phys.~Rev.~C {\bf 63},  024001  (2001).

\bibitem{deltuva:09d}
A. Deltuva, Phys.~Rev.~C {\bf 79},  054603  (2009).

\bibitem{gruebler:79a}
W. Gr\"uebler, R.~E. Brown, F.~D. Correll, R.~A. Hardekopf, N. Jarmie, and
  G.~G. Ohlsen, Nucl. Phys. {\bf A331},  61  (1979).

\bibitem{koersner:77}
I. Koersner, L. Glantz, A. Johansson, B. Sundqvist, H. Nakamura, and H. Noya,
  Nucl. Phys. {\bf A286},  431  (1977).

\bibitem{niessen:92}
P. Niessen, S. Lema\^itre, K.~R. Nyga, G. Rauprich, R. Reckenfelderb\"aumer, L.
  Sydow, H. Paetz~gen. Schieck, and P. Doleschall, Phys. Rev. C {\bf 45},  2570
   (1992).

\bibitem{kamada:00a}
H. Kamada, S. Oryu, and A. Nogga,
Phys. Rev. C {\bf 62},  034004   (2000).

\bibitem{thompson:11li}
I.~J. Thompson and M.~V. Zhukov, Phys. Rev. C {\bf 49},  1904  (1994).

\bibitem{nunes:96a}
F. Nunes, J. Christley, I. Thompson, R. Johnson, and V. Efros, Nucl. Phys. A
  {\bf 609},  43   (1996).

\bibitem{face:04}
I. Thompson, F. Nunes, and B. Danilin, Computer Physics Communications {\bf
  161},  87  (2004).

\bibitem{jensen:04a}
A. Jensen, K. Riisager, D. Fedorov, and E. Garrido, Reviews of Modern Physics
  {\bf 76},  215  (2004).

\bibitem{deltuva:17c}
A. Deltuva, Phys. Lett. B {\bf 772},  657  (2017).

\bibitem{hiyama:b19}
E. Hiyama, R. Lazauskas, J. Carbonell, and T. Frederico, Phys. Rev. C {\bf
  106},  064001  (2022).

\bibitem{o18-chin:25a}
N. Li, X.-R. Zhao, R. Zhang, L.-K. Wu, M.-J. Lyu, J.-J. He, W.-Y. Tong, Z.-Y.
  Zhang, and Y.-X. Li, Phys. Rev. C {\bf 112},  024004  (2025).

\bibitem{nndc}
https://www.nndc.bnl.gov/ensdf/  .

\bibitem{deltuva:13b}
A. Deltuva, Phys.~Rev.~C {\bf 87},  034609  (2013).

\bibitem{monteagudo:24a}
B. Monteagudo, F.~M. Marqu\'es, J. Gibelin, N.~A. Orr, A. Corsi, Y. Kubota, J.
  Casal, J. G\'omez-Camacho, G. Authelet, H. Baba, C. Caesar, D. Calvet, A.
  Delbart, M. Dozono, J. Feng, F. Flavigny, J.-M. Gheller, A. Giganon, A.
  Gillibert, K. Hasegawa, T. Isobe, Y. Kanaya, S. Kawakami, D. Kim, Y.
  Kiyokawa, M. Kobayashi, N. Kobayashi, T. Kobayashi, Y. Kondo, Z. Korkulu, S.
  Koyama, V. Lapoux, Y. Maeda, T. Motobayashi, T. Miyazaki, T. Nakamura, N.
  Nakatsuka, Y. Nishio, A. Obertelli, A. Ohkura, S. Ota, H. Otsu, T. Ozaki, V.
  Panin, S. Paschalis, E.~C. Pollacco, S. Reichert, J.-Y. Rousse, A.~T. Saito,
  S. Sakaguchi, M. Sako, C. Santamaria, M. Sasano, H. Sato, M. Shikata, Y.
  Shimizu, Y. Shindo, L. Stuhl, T. Sumikama, Y.~L. Sun, M. Tabata, Y. Togano,
  J. Tsubota, T. Uesaka, Z.~H. Yang, J. Yasuda, K. Yoneda, and J. Zenihiro,
  Phys. Rev. Lett. {\bf 132},  082501  (2024).

\end{thebibliography}

\end{document}